# Evolution of Photonic Time Stretch: From Analog to Digital Conversion to Blood Screening

**Bahram Jalali[1,2,3], Keisuke Goda[1,2], Ali Fard[1,2], and Sang Hyup Kim[1]**
[1]*Photonics Laboratory, Electrical Engineering Department, University of California, Los Angeles, CA 90095*
[2]*California NanoSystems Institute, University of California, Los Angeles, CA 90095*
[3]*Department of Surgery, David Geffen School of Medicine, University of California, Los Angeles, CA 90095*
*jalali@ee.ucla.edu*

**Abstract:** We show how the ability to slow down, amplify, and capture fast transient events can produce high-throughput real-time instruments ranging from digitizers to imaging flow cytometers for detection of rare diseased cells in blood.

**Introduction**

Outliers are rare occurrences of events that carry vital information. They appear in a wide range of systems such as communication links, sensors, and biological systems. To capture rare and non-repetitive evens, high-throughput instruments that operate in *real-time* are needed. The key feature required of the instrument is being able to record continuously over a long period of time, with not only fine temporal resolution, but also high throughput. In the context of optics, real-time instruments are needed to capture and quantify transient errors in communication systems and for screening of blood for pathogens and metastatic cancer cells.

The photonic time-stretch is a technique capable of slowing down, amplifying, and capturing fast events [1]. Its principles are those of a dispersive optical link in which group-velocity dispersion (GVD) is exploited to stretch an intensity- or phase-modulated chirped optical carrier [2]. Over the last 15 years, it has formed the basis of a class of fast real-time instruments that perform analog-to-digital conversion [1-2], spectroscopy [3], and imaging [4-6]. In the first step in a two-step process, the information to be captured is stamped onto the spectrum of a broadband optical pulse. This information can be a fast (wideband) electrical waveform such as that in a communication network, the spectroscopic signature of a material or process, or the image of an object. In the second step, the spectrum, and hence the information, is mapped into a time-domain serial data stream whose amplitude modulation is the information of interest. At the same time, the stream is optically slowed down by a sufficient amount so that it can be captured in real-time with an electronic digitizer. The second step is performed using GVD. This overview primarily describes its application to micrscopy.

**Time-Stretch Imaging**

Fueled by the innovative use of image processing or stimulated emission, a number of novel microscopy methods that break the diffraction limit of light have emerged in the last decade. However, the throughput of microscopy, currently limited by image-sensing technology, has not received much attention. An area where high-throughput imaging can make an important impact is label-free screening of complex biological fluids such as blood or other bodily liquids.

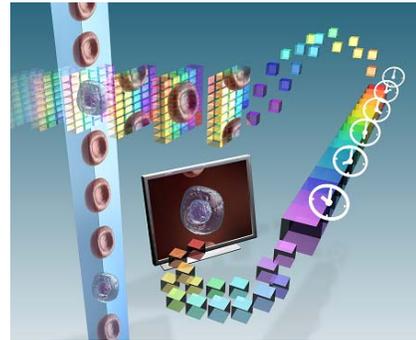

*Figure 1: Schematic of STEAM.*

Serial time-encoded amplified microscopy (STEAM) [4,5] maps a multi-dimensional image into a 1D serial time-domain optical waveform that is stretched in time, optically amplified, and converted to an electronic waveform, and then captured with a real-time digitizer (Figure 1). This is done by first encoding the spatial coordinates of the object onto the spectrum of a broadband pulse with a 1D (line scan) or 2D rainbow created by a spatial disperser. In our original demonstration [4,5], the optical source was a mode-locked femtosecond fiber laser with 1590 nm center wavelength, 15 nm bandwidth, and pulse repetition rate of 37 MHz. For 1D imaging, a diffraction grating or prism is used as a spatial disperser while for 2D imaging, a pair consisting of a diffraction grating and an orthogonally oriented virtually-imaged phased array is employed. The image encoding occurs when the spatially-dispersed pulse or rainbow is reflected or transmitted through the object, after which it returns to the spatial disperser where the different frequency components of the pulse are recombined. An optical circulator directs the pulse into a time-stretch Fourier transformer with internal Raman amplification. This is realized using a Raman-pumped dispersive fiber in which the image pixels are serialized into a 1D data stream as they experience different group delays. The multi-dimensional spatial image now resembles serial data in a communication link, and is captured with a real-time oscilloscope. Optical image amplification allows STEAM to operate at shutter speeds of tens of picoseconds and frame rates of tens of millions of frames per second and with very low illumination power (only a few mW). In contrast, ultrafast CMOS cameras have a shutter speed of about 300ns and frame rate of less than 1 MHz, and due to lack of optical image amplification, they require much higher illumination power. For biological applications where the illumination is focused onto a microscopic field of view, this can lead to damage or destruction of the sample being studied. Animated movies that illustrate the functionality of STEAM can be viewed at Ref. [7]. It can also operate in the phase-contrast mode for 3D imaging. The performance of STEAM, as it relates to its optical resolution and how it depends on the specifications of the optical and electronic components, is described in Ref. [8]. STEAM has been used to observe the dynamics of laser ablation and microfluidic flow [4].

As a further proof of STEAM's capability for screening of complex fluids, it has been used to monitor ultrafast microfluidic flow (Figure 2) [4]. Water-suspended microspheres with diameters similar to blood cells (5 – 30 μm) were pumped into a hollow fiber with a diameter of 50 μm (Figure 2a). STEAM captured thousands of frames in real time (limited only by the memory of the real-time oscilloscope). Figure 2b shows one out of every seventeen snapshots for clarity, whereas Figure 2c shows consecutive frames with the frame resolution of 163 ns. The flow of metal microspheres from right to left is clearly observed and shows a velocity of 2.4 m/s commensurate with the state-of-the-art flow cytometers that operate at 1 – 5 m/s. Due to their limited frame rate and shutter speed, conventional cameras (i.e., CCD and CMOS cameras) are incapable of imaging such a high-speed flow. This is the first time that such ultrafast microfluidic flow has been observed in real time with such a fine temporal resolution. An emerging and vital application that is being pursued is the detection of circulating tumor cells – the vanishingly rare tumor cells that circulate in a patient's blood and are forerunners of the metastatic stage responsible for 90% of cancer mortalities [9].

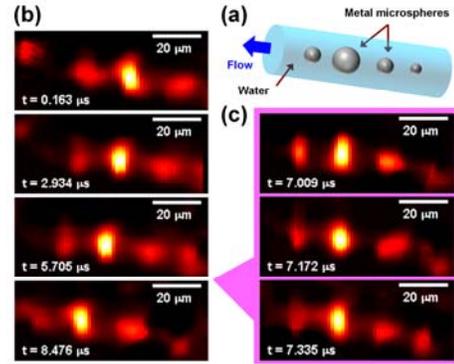

*Figure 2: Real-time observation of ultrafast microfluidic flow with STEAM.*

High-throughput imaging also requires fast image processing and the speed of the STEAM imager is significantly beyond the speed of electronic processors. The imager's ability to perform optical image serialization and amplification naturally lends itself to optical analog image correlation detection *in the time domain* [10]. Figure 3 shows real-time image recognition of binary image patterns (i.e., barcodes) at a frame rate of 37 MHz. In these experiments, the reference pattern that is synchronized with the image-encoded pulse stream modulates the serialized pixel stream in an electro-optic modulator. The processed image is optically integrated in time by compressing it in a dispersive fiber before photodetection. The complementary output of the modulator is used to generate the complementary correlation output. The difference of the positive and negative outputs is computed in the digital domain to show the final correlation output, indicating whether the test pattern is matched with the reference.

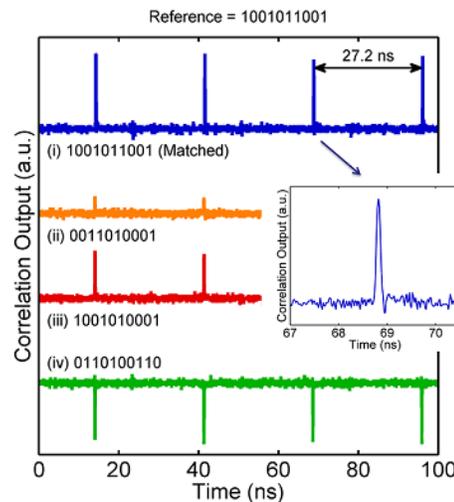

*Figure 3: Correlation output for various test binary image patterns against a fixed reference pattern.*

In summary, the ability to manipulate the time scale of fast transient signals enables a new imaging modality with an unprecedented throughput. This talk describes the technology and its various applications.

1. F. Coppinger, A.S. Bhushan, and B. Jalali, "Time magnification of electrical signals using chirped optical pulses," *Electronics Letters* 34, 399 (1998)
2. Y. Han and B. Jalali, "Photonic time-stretched analog-to-digital converter: fundamental concepts and practical considerations," *Journal of Lightwave Technology* 21, 3085 (2003)
3. P. Kelkar, F. Coppinger, A. S. Bhushan, and B. Jalali, "Time-domain optical sensing," *Electronics Letters* 35, 1661 (1999)
4. K. Goda, K. K. Tsia, and B. Jalali, "Serial time-encoded amplified imaging for real-time observation of fast dynamic phenomena," *Nature* 458, 1145 (2009)
5. Bahram Jalali, Patrick Soon-Shiong, Keisuke Goda, "Breaking Speed and Sensitivity Limits Real-Time Diagnostics with Serial-Time Encoded Amplified Microscopy," Optik & Photonik, No. 2, 32 (2010)
6. Bahram Jalali, Daniel R. Solli, Keisuke Goda, Kevin Tsia, Claus Ropers, "Real-time Time-stretch Instruments, Rogue Events, and Photon Economics" The European Physical Journal Special Topics. Vol. 185 (July 2010)
7. http://www.youtube.com/watch?v=a6nVy-fiNj8
8. K. K. Tsia, K. Goda, D. Capewell, and B. Jalali, "Performance of serial time-encoded amplified microscope," *Optics Express* 18, 10016 (2010).

9. J. Kaiser, "Cancer's circulation problem," *Science* 327, 1072 (2010)
10. S. H. Kim, K. Goda, A. Fard, and B. Jalali, "An optical time-domain analog pattern correlator for high-speed real-time image recognition," *Optics Letters* (2010) in press